\documentclass[aps,prl,floatfix,twocolumn,superscriptaddress,showpacs]{revtex4}

\pdfoutput=1
\usepackage{graphicx}
\usepackage{amssymb}
\usepackage{amsmath}

\newcommand {\ket}[1] {|#1 \rangle}
\newcommand {\bra}[1] {\langle#1 |}
\newcommand {\braket}[2] {\langle #1 | #2 \rangle}

\newcommand{\lr}[1]{\left( #1 \right)}
\newcommand{\comm}[2]{\left[#1,#2\right]}
\newcommand{\mean}[1]{\langle #1 \rangle}
\newcommand{\f}[1]{\hat{\psi}(#1)}
\newcommand{\fd}[1]{\hat{\psi}^{\dag}(#1)}
\newcommand{\F}[2]{\hat{\psi}_{#2}(#1)}
\newcommand{\Fd}[2]{\hat{\psi}^{\dag}_{#2}(#1)}
\newcommand{\no}{\nonumber}
\begin{document}

\title{Dressed, noise- or disorder- resilient optical lattices}

\author{Hannes Pichler} \affiliation{Institute for Theoretical Physics,
  University of Innsbruck, A-6020 Innsbruck, Austria}\affiliation{Institute
  for Quantum Optics and Quantum Information of the Austrian Academy
  of Sciences, A-6020 Innsbruck, Austria}

\author{Johannes Schachenmayer} \affiliation{Department of Physics and Astronomy, University of Pittsburgh, Pittsburgh, Pennsylvania 15260, USA}
\affiliation{Institute for Theoretical Physics,
  University of Innsbruck, A-6020 Innsbruck, Austria}\affiliation{Institute
  for Quantum Optics and Quantum Information of the Austrian Academy
  of Sciences, A-6020 Innsbruck, Austria}

\author{Jonathan Simon}
\affiliation{Department of Physics, Harvard University, 17 Oxford Street, Cambridge, Massachusetts 02138, USA}

\author{Peter~Zoller} \affiliation{Institute for Theoretical Physics,
  University of Innsbruck, A-6020 Innsbruck, Austria}\affiliation{Institute
  for Quantum Optics and Quantum Information of the Austrian Academy
  of Sciences, A-6020 Innsbruck, Austria}

\author{Andrew~J.~Daley} \affiliation{Department of Physics and Astronomy, University of Pittsburgh, Pittsburgh, Pennsylvania 15260, USA}

\date{May 28, 2012}


\begin{abstract}
External noise is inherent in any quantum system, and can have especially strong effects for systems exhibiting sensitive many-body phenomena. We show how a dressed lattice scheme can provide control over certain types of noise for atomic quantum gases in the lowest band of an optical lattice, removing the effects of lattice amplitude noise to first order for particular choices of the dressing field parameters. We investigate the non-equilibrium many-body dynamics for bosons and fermions induced by noise away from this parameter regime, and also show how the same technique can be used to reduce spatial disorder in projected lattice potentials.
\end{abstract}

\maketitle

Quantum many-body systems are typically very sensitive to external noise sources and disorder, especially when they exhibit complex many-body phenomena that are protected by small energy gaps. 
Experiments with quantum gases in optical lattices \cite{Blo08,Lew07,toolbox} are rapidly approaching such regimes. A particularly relevant example is the quest to realize quantum magnetism \cite{Sch08,Joe08,simulator,Tro09}, where the relevant energy scales are set by the super-exchange interaction \cite{Jor10,Wel09,Med10,McK10,McK11}, and are typically very small (of the order of tens to hundreds of Hertz). In these systems, dissipative mechanisms such as spontaneous emissions \cite{Pic10} lead to a form of \emph{quantum} noise, and \emph{classical} noise enters through fluctuations of the lattice potential. In order to work in these regimes in experiments, it is then imperative to understand how this noise can be controlled, e.g., by designing the system to be resilient to certain types of noise and disorder. We must also understand the detailed non-equilibrium many-body dynamics generated by classical and quantum noise in such systems, and determine how the resulting heating processes depend on the many-body states that are present. Here we analyze the dynamics of atoms in the lowest band of an optical lattice arising from low-frequency intensity fluctuations of the lattice, and design a dressed lattice setup that is resilient to this type of noise. We solve the stochastic differential equations (SDE) for heating of either a single species of bosons initially in Mott Insulator or superfluid ground states, or two-species fermions near an anti-ferromagnetic ground state, and demonstrate that the noise resilience is robust in parameter space. Moreover, we show that this scheme can be used to flatten disorder potentials in experiments with projected lattice potentials \cite{Bak09,Sim11}, which are a key potential route towards homogeneous lattice systems. 

\begin{figure}[tb]
\includegraphics[width=8.5cm]{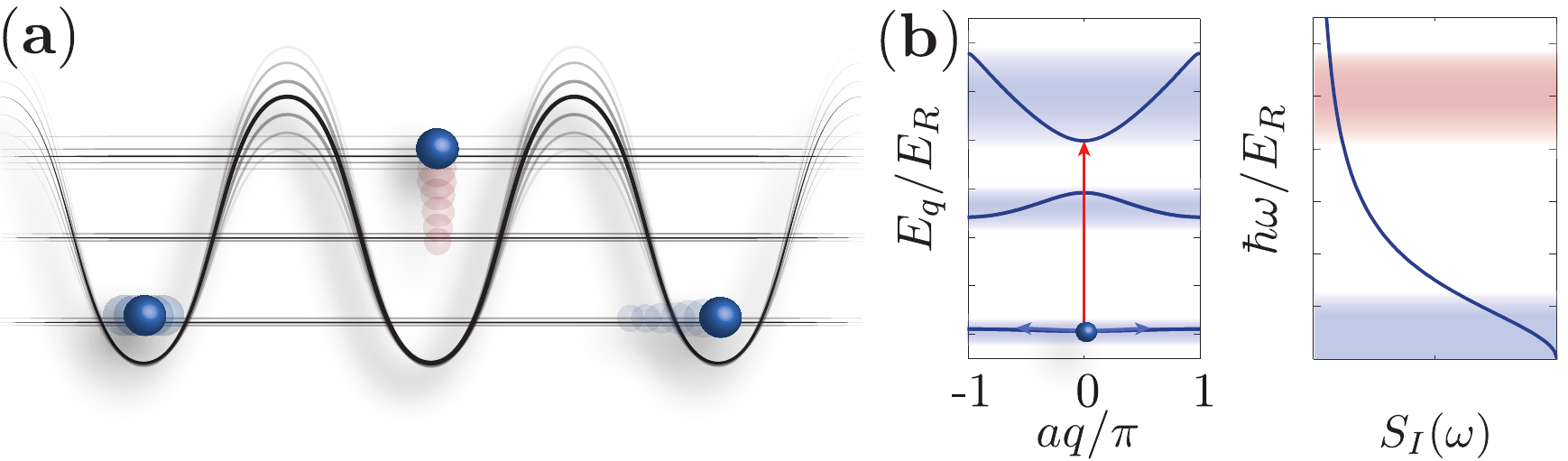}
\caption{(color online) (a) Laser intensity fluctuations give rise to amplitude noise on an optical lattice potential which can give rise to different heating processes depending on the intensity noise spectrum $S_I(\omega)$. (b) Noise with frequencies of the order of Bloch band separations can give rise to \emph{inter-}band transitions (red arrow), while processes with frequencies of the order of $J$ and $U$ will give rise to \emph{intra-}band heating (blue arrows).} \label{fig:Fig1}
\end{figure}

We begin with the example of cold, dilute bosons in an optical potential $V_0(\vec{x})+\delta V(\vec{x},t)$, where the fluctuations $\delta V(\vec{x},t)$ represent either (i) multiplicative time-dependent noise ($\delta V(\vec{x},t) \equiv V_{\rm noise}(\vec{x}) \delta V(t)$), or (ii) static disorder ($\delta V(\vec{x},t)\equiv \delta V_{\rm dis}(\vec{x})$). The corresponding dynamics are described by the Hamiltonian ($\hbar\equiv 1$)
\begin{eqnarray}
H_{B}&=&\int {\rm d}^3 x \hat \psi^\dag (\vec{r})\left [-\frac{\nabla^2}{2m}  + V_0(\vec{x}) + \delta V(\vec{x},t) \right] \hat \psi (\vec{x}) \nonumber \\
& &+ \int {\rm d}^3 x {\rm d}^3 x' \hat \psi^\dag (\vec{r}) \hat \psi^\dag (\vec{x}') U(\vec{x}-\vec{x}')  \hat \psi (\vec{x}') \hat \psi (\vec{x}),\nonumber 
\end{eqnarray}
where $\hat \psi(\vec{x})$ is a bosonic field operator, and $U(\vec{x}-\vec{x}')$ specifies the two-body interactions. Note that the generalization to multiple species and to fermions is straight-forward. For noise, the corresponding dynamics are described by a multiplicative stochastic differential equation (SDE) \cite{Klo_book,Gar_book}, which arise for single-particle systems in quantum optics, and often have non-perturbative solutions. Here we now investigate an example of such dynamics in our many-body system. 

We aim to identify regimes in which the system becomes insensitive to certain types of noise and disorder. In particular, for classical noise we can write the many-body Hamiltonian $H_{B}=H_0+\delta V(t) H_1$, where $H_0$ is the Hamiltonian without the noise component, and $\delta V(t)$ specifies the time-dependence of fluctuations, which are proportional to $H_1$. Below we will show that for some types of noise, we can engineer a dressed potential such that $[H_0, H_1]=0$, and the stochastic term can be reinterpreted as a noise on the time parameter of the Schr\"odinger equation. The system will then be resilient against the noise, e.g., with eigenstates remaining stationary in time and the mean energy of the system remaining constant.

\emph{Amplitude noise in an optical lattice:} We now specialize to the case of atoms in the lowest Bloch band of an optical lattice, in the presence of amplitude noise on the lattice depth $V=V_0+\delta V(t)$, arising from fluctuations of the laser intensity. In experiments, this will be associated with a particular noise spectrum, depending on the technical details of the setup \cite{noiseexample}. As depicted in Fig.~1, different components of this noise spectrum will give rise to different dynamical processes. While noise at frequencies of the order of the band separation can give rise to \emph{inter}-band processes in which particles are transferred to higher Bloch bands, noise at lower frequencies of the order of the tunneling parameter $J$ and on-site interaction strength $U$ will give rise to \emph{intra}-band heating for atoms within the lowest band. We note that for relevant frequency scales, intensity fluctuations on the lattice beams give rise to \emph{global} noise on the lattice potential (as the fluctuations are much slower than the time for light to propagate across the system).

In the case where the noise is weak for inter-band processes (e.g., the spectrum is dominated by $1/f$ low-frequency noise), the evolution of atoms in the lowest Bloch band is governed to first order by a stochastic model for intra-band heating (see the supplementary material for more details of the derivation),
\begin{align}
i\frac{d\ket{\psi}}{dt}=\ket{\psi} \left[ H\lr{J,U}+H\lr{\frac{dJ}{dV},\frac{dU}{dV}} \delta V(t)\right]\ket{\psi}, \label{sde}
\end{align}
where for bosons $H(J,U)$ is the Bose-Hubbard model
\begin{align}
H(J,U)=-J \sum_{\langle {i},{j}\rangle}b_{{i}}^{\dag}b_{{j}}+\frac{U}{2}\sum_{{i}} b_{{i}}^{\dag}b_{{i}}^{\dag}b_{{i}}b_{{i}} + \sum_i \varepsilon_i b_i^\dag b_i. \nonumber
\end{align}
Here, $b_i$ is the bosonic annihilation operator for an atom on site $i$, $J$ is the tunnelling rate, and $U$ is the on-site interaction energy shift for two atoms. The external trapping potential is given by $\varepsilon_i$, and we will initially set $\varepsilon_i=0$, before returning to the trapped case below. 

In a typical experimental setup, we have $dJ/dV<0$ and $dU/dV>0$, as an increase in the lattice depth increases tunnel barriers and confines the atoms more tightly on each site \cite{toolbox}, so that the noise is anti-correlated on $J$ and $U$. Below we will present a dressed lattice scheme where the relative noise on $J$ and $U$ can be controlled so that the noise is correlated, and
\begin{align}
\xi \equiv \frac{1}{J}\frac{dJ}{dV}-\frac{1}{U}\frac{dU}{dV}=0, \label{condition}
\end{align}
implying that the relative change in $J$ and $U$ is identical, $d(U/J)/dV=0$.
This defines a \emph{parameter space of sweet spots}, where the stationary $H_0=H(J,U)$ and time-varying $H_1=H(d J/d V, d U/d V)$ Hamilitonians commute $[H_0,H_1]=0$, so that the system is resilient against noise. 

\textit{Non-equilibrium stochastic dynamics and heating for correlated and anti-correlated noise:} Let us first consider the many-body dynamics away from $\xi=0$. In the limit of white noise on $V$,  $\langle \delta V(t)\rangle=0$, $\langle \delta V(t) \delta V(t') \rangle=S \delta(t-t')$, eq.\eqref{sde} becomes a Stratonovich SDE \cite{Klo_book,Gar_book}, and we can compute the mean energy increase in the system,
\begin{align}
  \mean{\dot
    H}&=\frac{S}{2}\lr{\frac{1}{J}\frac{dJ}{dV}-\frac{1}{U}\frac{dU}{dV}}^2\mean{\comm{\comm{H_J}{H_U}}{H_J}},\label{eq:Global10}
\end{align}
where $H_J=-J\sum_{\langle i,j \rangle} b_i^\dag b_j$ and $H_U=(U/2)\sum_i b_i^\dag b_i^\dag b_i b_i$ denote the
kinetic and interaction energy terms in the Bose-Hubbard model. We therefore see that the mean rate of energy increase in this limit grows as $\xi^2$ away from the sweet spots, and is proportional to the number of particles $N$ (as the commutators are local in space).

\begin{figure}[tb]
\includegraphics[width=8.5cm]{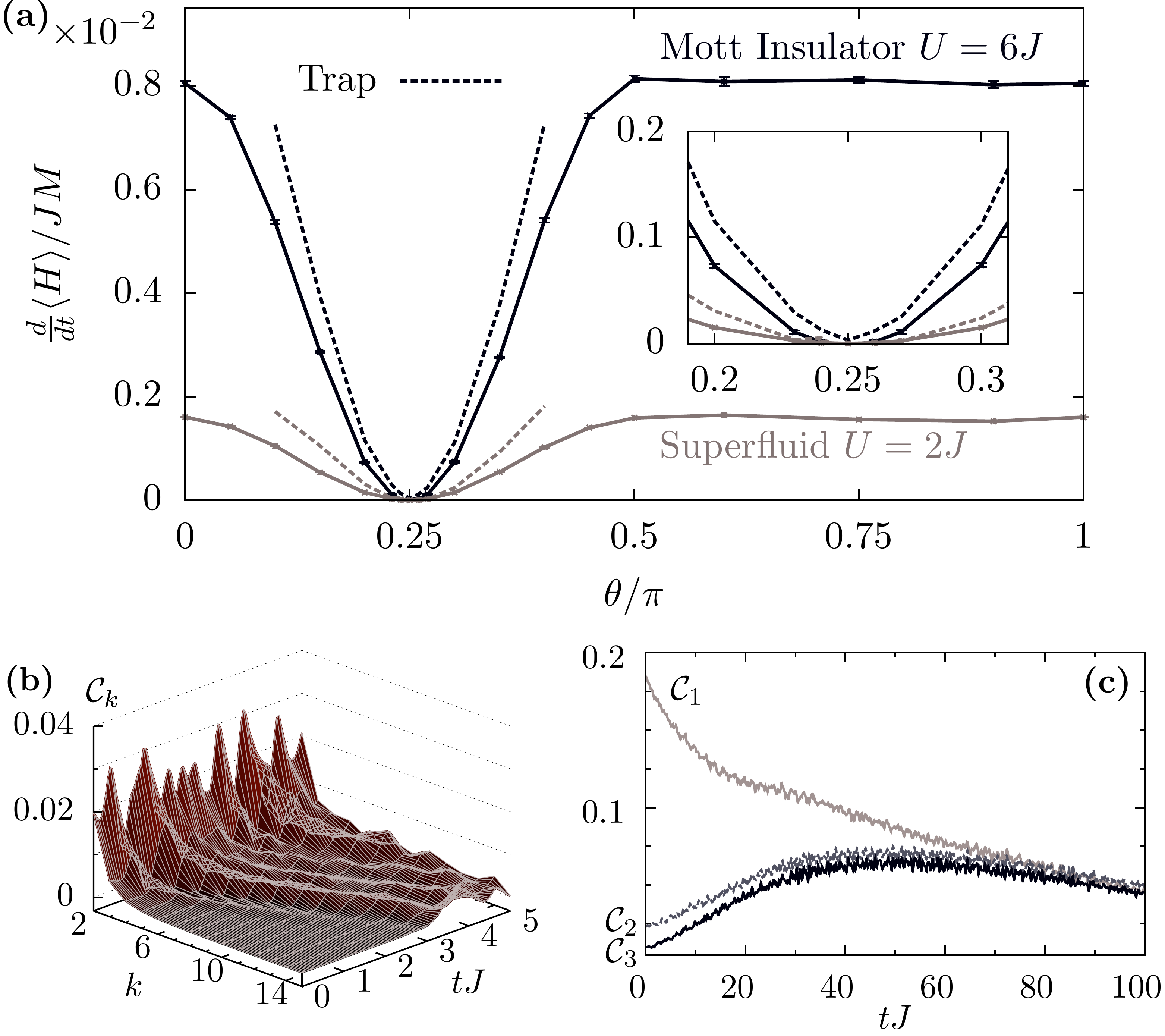}
\caption{(a) Short-time ($tJ\leq 2$) heating rates of superfluid ($U=2J$) and Mott insulator states ($U=6J$) in 1D as a function of the relative magnitude of noise on $J$ and $U$, $\theta$. Results are from linear regression over $500$ t-DMRG trajectories in a system with $M=30$ sites. In both cases heating is strongly suppressed in the vicinity of the sweet spot at $\theta \sim 0.25 \pi$ (expanded in the inset). Dashed lines show results in the presence of a harmonic trap with $\varepsilon_{i=0}-\varepsilon_{i=M/2} = 8 J$, $\sqrt{S} (d\varepsilon_i/dV)/\varepsilon_i = 5 \times 10^{-3}J^{-1/2}$. (t-DMRG bond dimension $D=100$). (b, c) The effect of amplitude noise on parity-parity correlations on an initial Mott insulator state ($U=6J$), with anti-correlated noise $\theta=0.75\pi$. (b) Short-time evolution for a single noise trajectory with $M=30$ (t-DMRG bond dimension $D=200$). (c) Long-time evolution of these correlations calculated via exact diagonalization in a small system with $M=10$ sites averaged over $1000$ noise trajectories.  (For all parts, $\lambda=0.02J^{-1/2}$, and timestep $\Delta t = 10^{-2}/J$.)} \label{fig:Fig2}
\end{figure}
\begin{figure}[tb]
\includegraphics[width=8cm]{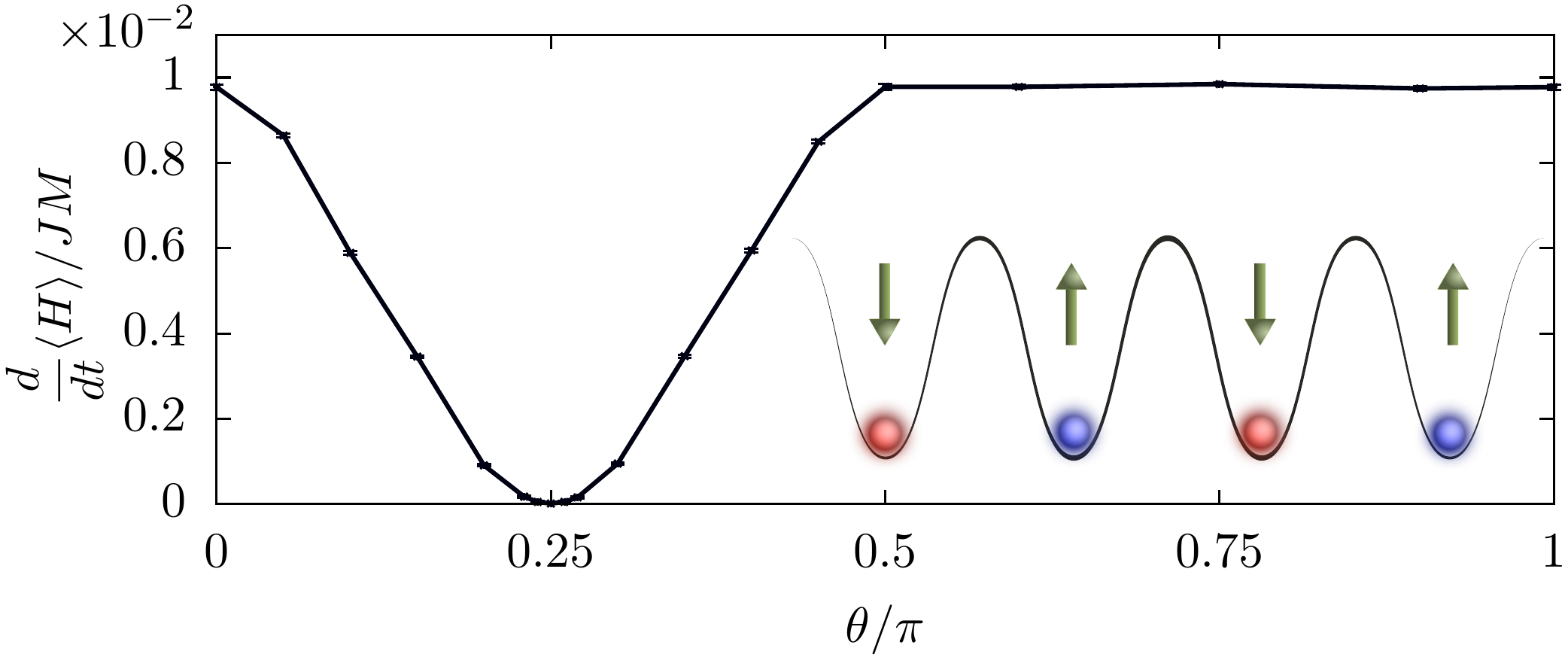}
\caption{Heating rates around the sweet spot for a two-species Fermi-Hubbard model in 1D, with an initial anti-ferromagnetic ground state with inter-species interaction $U=10J$. The noise is parametrized as in Fig.~2.  (t-DMRG bond dimension $D=100$, timestep size $\Delta t = 10^{-2}/J$, $500$ noise trajectories).} \label{fig:Fig3}
\end{figure}

It is possible to study many-body dynamics in the presence of noise for varying noise statistics and correlations. We take the example of white noise and propagate eq.~\eqref{sde} as a many-body SDE using t-DMRG methods for a 1D system \cite{Vidal04,tdmrg1,tdmrg2,dmrgrev}, sampling over noise realizations. We parameterize the correlations between the noise on $J$ and $U$ by $\theta$ and $\lambda$ as $ \sqrt{S}(dU/dV)/U= \lambda \sin^2(\theta)$ and $\sqrt{S}(dJ/dV)/J= \lambda \cos^2(\theta)$ for $0 \leq \theta < \pi/2 $; $\sqrt{S}(dJ/dV)/J=- \lambda \cos^2(\theta)$ for $\pi/2 \leq \theta < \pi$. The usual anti-correlated case corresponds to $\theta>\pi/2$, and the sweet spot $\xi=0$ of eq.~\eqref{condition} to $\theta=\pi/4$. In Fig.~\ref{fig:Fig2}a we plot the rate of energy increase starting in a Mott Insulator (MI) or superfluid (SF) ground state as a function of $\theta$, keeping the sum of the relative noise on $J$ and $U$ terms constant. In agreement with eq.~\eqref{eq:Global10}, we observe that for anti-correlated noise ($\theta>\pi/2$), the rate of energy increase depends only on $S$, and not on $\theta$, whereas for correlated noise ($\theta<\pi/2$) we observe a quadratic increase in the heating rate around the sweet spot. The effects of classical noise are significant for both MI and SF states. Note that in an experiment, this characteristic would help in distinguishing heating due to amplitude noise from heating due to spontaneous emissions, which have very weak influence on Mott Insulator states for atoms remaining in the lowest band \cite{Pic10}. 

In Fig.~2a we also show the heating in the presence of a harmonic trapping potential. When such a trapping potential is included in $H_0$, it is no longer possible to fulfill the condition $[H_0,H_1]=0$ exactly. However, for typical values of the trapping frequencies the residual heating is extremely small at $\xi=0$, and in Fig.~2a is over two order of magnitude smaller than for anti-correlated noise.

We can also ask how the many-body state changes as a result of the heating. For the Mott Insulator state, the noise produces correlated particle-hole pairs that spread through the system as a function of time. This is shown in Fig.~\ref{fig:Fig2}b, where we plot parity-parity correlation functions $\mathcal C_k(t)=\langle \hat s_l \hat s_{l+k} \rangle -\langle \hat s_l \rangle \langle s_{l+k} \rangle$ with $\hat s_l=\exp[i\pi( \hat n_l - 1)]$, which can be measured in experiments with a quantum gas microscope \cite{Che12,Bar12}. Initially, the nearest-neighbor parity-parity correlations are strongest, resulting from virtual tunneling of particles to neighboring sites with an amplitude $J/U$. The amplitude noise produces real particle-hole excitations, which transfer initially to next-neighbor sites and then spread through the system, while the nearest neighbor correlation functions decrease monotonically, as shown in Fig.~\ref{fig:Fig2}c. Deep in the Mott insulator limit ($U/J\sim 40$), we also see an increase in off-diagonal correlations $\langle b_i^\dag b_j \rangle$ for a fixed noise realization, that corresponds to an increased condensate fraction for a finite system (computed as the largest eigenvalue of the single-particle density matrix). For superfluid states, the heating takes on a different form. In a weakly interacting gas, we can trace the heating to creation of pairs of Bogoliubov excitations, which always leads to a decrease in the condensate fraction \cite{Pic12}, and energy increase per particle at a rate $\dot E/N\approx S\xi^2zJU^2 \bar n $, where $z$ is the number of nearest neighbors  and $\bar n$ is the filling factor.

We observe similar behavior for a two-species Fermi Hubbard model,
 \begin{align}
H_{FH}=-J \sum_{\langle {i},{j}\rangle}c_{{i,\sigma}}^{\dag}c_{{j,\sigma}}+U\sum_{{i}} c_{{i\uparrow}}^{\dag}c_{{i\uparrow}}c_{{i,\downarrow}}^{\dag}c_{{i,\downarrow}} 
\end{align}
where $c_{i,\sigma}$ is a fermionic annihilation operator for a particle in state $\sigma \in \{\uparrow,\downarrow \}$ on site $i$. In Fig.~3 we plot a similar analysis to Fig.~2a, beginning in the ground state for $U\ll J$ at half filling, where the state exhibits antiferromagnetic order \cite{Jor10} (which in 1D is characterized by algebraically decaying anti-ferromagnetic correlations). The heating mechanism here involves excitations above the underlying Mott Insulator state, and we again see that the noise here is also robustly suppressed around the sweet spot, as in the Bosonic case. In a mean-field approximation, the heating rate per particle is given by $\dot E/N\approx S\xi^2 UzJ^2$.

\begin{figure}[tb]
\includegraphics[width=8cm]{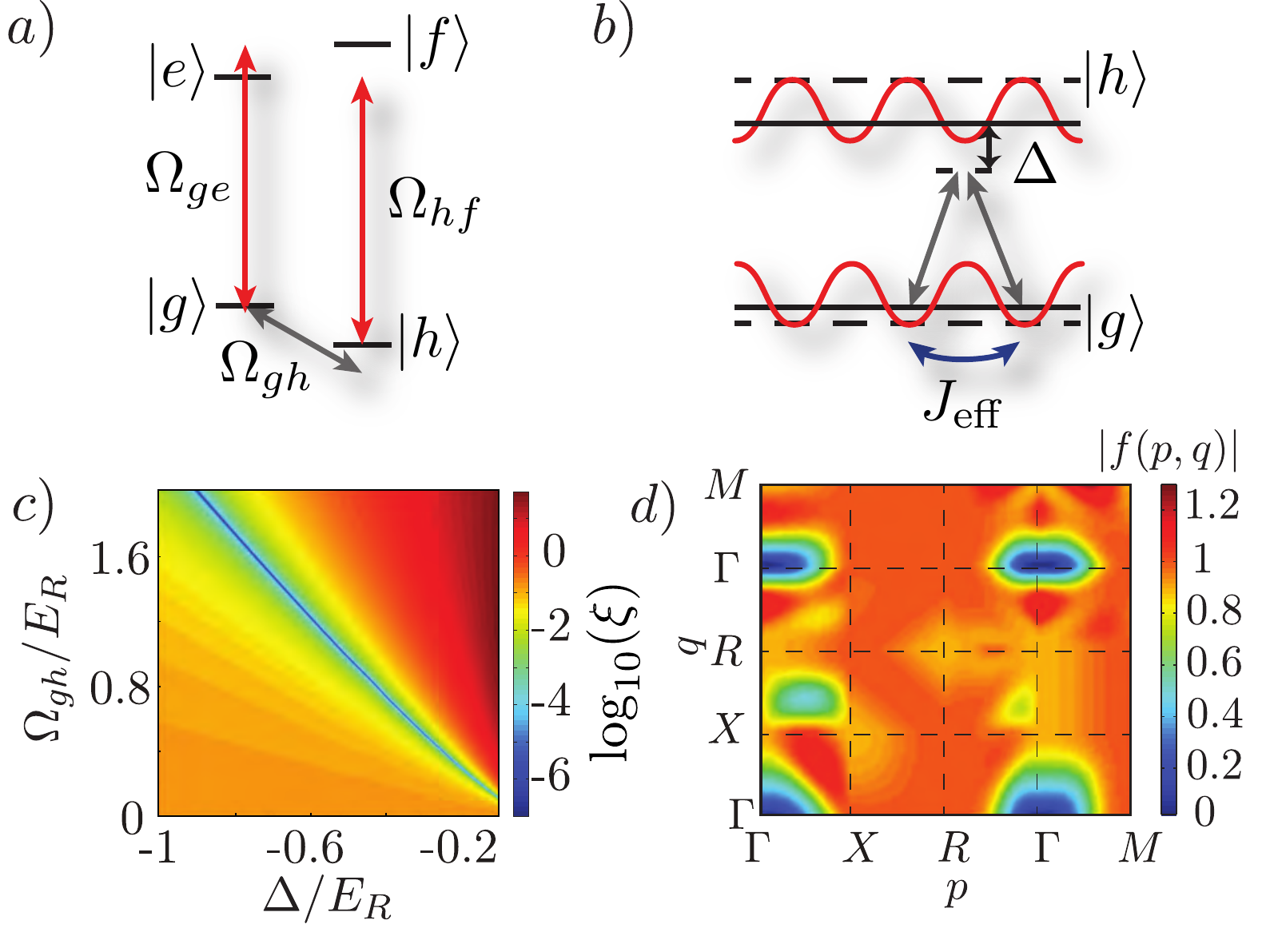}
\caption{(a) Schematic plot showing coupled internal states used to create dressed lattices for noise or disorder suppression. Two long-lived states $\ket{g}$ and $\ket{h}$ have far-detuned optical lattices created from the same laser (i.e., with identical intensity fluctuations). (b) These internal states are coupled to give rise to a dressed lattice. If the lattice for $\ket{g}$ is blue-detuned, and the lattice for $\ket{h}$ red-detuned (as indicated by the dashed lines, showing the energy at zero AC-Stark shift), then when the lattice depth increases, the effective detuning decreases, allowing for a larger admixture of $\ket{h}$, and hence an increase in the effective tunnelling rate of the dressed atoms. (c) Correlation parameter $\xi$ for different detunings and couplings in an isotropic 3D lattice of depth $V=7E_R$ along each dimension. (d) Reduction of disorder, shown as the absolute value of the multiplication factor $|f(p,q)|$ from eq.~\eqref{disorder} for a 3D lattice with $V=7E_R$, plotted along the lines of symmetry of the first Brillouin Zone, i.e., connecting the points $\Gamma=(0,0,0), X=(\pi/a,0,0), R=(\pi/a,\pi/a,\pi/a)$, and $M=(0, \pi/a, \pi/a)$, computed from coupling the lowest Bloch bands.} \label{fig:Fig4}
\end{figure}

\textit{Dressed lattice setup to engineer sweet spots:} A correlated noise regime with $\xi=0$ can be obtained with the dressing scheme depicted in Figs.~\ref{fig:Fig4}a,b. We consider two internal atomic states, a primary state $\ket{g}$, which is trapped in a blue-detuned optical lattice, and an auxilliary state $\ket{h}$, which is trapped in a red-detuned optical lattice produced by the same laser. This can be achieved, e.g., by tuning the laser in the middle of the fine-structure splitting of an alkali atom \cite{spindep1,spindep2}, or by using anti-magic wavelength lattices for alkaline earth atoms \cite{AEatom1,yi08,aeshort}. The states $\ket{g}$ and $\ket{h}$ are then coupled to produce the dressed lattice. As shown in Fig.~\ref{fig:Fig4}b, this coupling (with coupling constant $\Omega_{gh}$ and  detuning $\Delta$) effectively gives rise to an additional tunnelling mechanism for the $\ket{g}$ atoms to move between sites, via a virtual coupling to the state $\ket{h}$. A small increase in lattice depth will then shift the effective detuning of this coupling, because the energy levels in the red-detuned lattice shift $\propto -(V+\delta V)$, whereas those in the blue-detuned lattice shift $\propto \sqrt{V+\delta V}$. Thus, if the detuning is chosen appropriately, a small increase in lattice depth can lead to a decrease in the effective detuning, and hence an increase in tunneling due to coupling via the state $\ket{h}$. For an appropriate parameter choice, this will more than compensate for the decrease in bare tunnelling for $\ket{g}$, so that $dJ/dV>0$. 

As detailed in the supplementary material, the full Hamiltonian is still periodic with periodicity $a=\pi/k$, and we can calculate the associated bandstructure exactly. This allows us to compute Wannier functions, and define effective parameters $U_{\rm eff}$ and $J_{\rm eff}$ for the Hubbard model in the usual way \cite{toolbox}, as well as determining their dependence on the lattice depth. For typical lattice depths, we can always find realistic values for the parameters $\Omega_{gh}$ and $\Delta$ such that the condition of eq.~\ref{condition} are satisfied. In Fig.~\ref{fig:Fig4}c, we give an example for a lattice depth $V=7E_R$ (with $E_R=\hbar^2k^2/2m$), where we show a plot of $\xi$ as a function of $\Omega_{gh}$ and $\Delta$. We see a line of choices for these parameters where the condition $\xi=0$ is exactly fulfilled. More details, as well as approximate analytical values for $\Omega_{gh}$ and $\Delta$ from perturbation theory calculations are provided in the supplementary material.

We note that we require $U_{\rm eff}<\Omega_{gh},\Delta$, but that $\Omega_{gh},\Delta$ can be comparable to the energy gap between Bloch bands $\omega$ in this scheme. This is important, both to fulfill the condition on $U_{\rm eff}$, and because a larger $\Delta$ means that the scheme will be more robust, e.g., against magnetic field fluctuations that could shift the effective detuning as the energies of $\ket{h}$ and $\ket{g}$ shift. One could consider engineering a similar scenario for non-interacting particles by coupling a single internal state to a higher Bloch band of the lattice potential. However, there the condition $U_{\rm eff}\ll\Delta\ll \omega$ is required and is very difficult to fulfill. We also note that we require a low-noise driving field for the coupling, which can be provided either by an RF generator for alkali atoms, or a clock laser for alkaline earth atoms \cite{AEatom1}.

\textit{Removal of spatial disorder:} An alternative application of the dressed lattices is found if we choose $\Delta$ and $\Omega_{gh}$ such that ${d\varepsilon_{\rm{eff}}}/{dV}=0$. This condition can flatten spatial disorder potentials, especially as arises in projected lattices \cite{Bak09,Sim11}. However, it can be used to decrease any slowly-varying, shallow potential. If we transform the Hamiltonian into a quasimomentum representation, then by considering the coupling between the lowest bands in the lattice for $\ket{g}$ and $\ket{h}$ we can show that the disorder hamiltonian $H_D=\sum_{p,q} \delta \varepsilon_g(q)  a_{p+q,g}^{\dag} a_{p,g}$ (where $a_{p,\sigma}$ is a bosonic annihilation operator for a particle with internal state $\sigma$ and quasimomentum $p$) becomes an effective disorder in the dressed lattice 
\begin{equation}
H_D\rightarrow\sum_{p,q} \delta \varepsilon_g(q)f(p,q)a_{p+q,-}^{\dag}a_{p,-}, \label{disorder}
\end{equation}
with $\sigma=-$ indicating the lower energy dressed state. That is, the ``disorder'' is multiplied by the factor $f(p,q)$. In Fig.\ref{fig:Fig4} we show $|f(p,q)|$ computed for an isotropic 3D lattice, and plotted along the lines of symmetry in the Brillouin zone. We see that this technique works very well either for disorder that varies slowly on the scale of a lattice site, i.e., $|q|$ is not too large, or if the quasimomentum of the state $|p|$ is not too large. For disorder of the scale of two sites, there is already a reduction of the disorder by a factor greater than 2, and for the small quasimomentum states that are typically most affected by the disorder potential, the reduction can be more than an order of magnitude.

\textit{Summary and outlook:} We have presented a scheme for dressed lattices that provide both a parameter regime where amplitude noise is eliminated for atoms in the lowest band of the optical lattice, and another regime where disorder arising from spatial intensity fluctuations can be suppressed. This scheme works equally well for bosons and fermions, and could be used for studies of non-equilibrium many-body dynamics driven by noise.

\emph{Acknowledgements:} We thank S.~Blatt, M.~Lukin, and the groups of Immanuel Bloch and Wolfgang Ketterle for helpful and motivating discussions. This work was supported in part by the Austrian Science Fund through SFB F40 FOQUS, and by a grant from the US Army Research Office with funding from the DARPA OLE program. Computational resources were provided by the Center for Simulation and Modeling at the University of Pittsburgh.

¥

\newpage

\section*{SUPPLEMENTARY MATERIAL}
\section{Derivation of the lowest band stochastic Hubbard model}
The starting point for the derivation is the many-body Hamiltonian ($\hbar\equiv 1$)
\begin{align}
H&=\int d^3x\, \fd{\vec{x}}\lr{-\frac{\nabla^2}{2m} +V_{\rm opt} (\vec{x},t)}\f{\vec{x}}\no\\
&+\frac{g}{2}\int d^3x\, \fd{\vec{x}}\fd{\vec{x}}\f{\vec{x}}\f{\vec{x}}.
\end{align}
For notational simplicity we present here only the one dimensional case, but this can be immediately generalized to higher dimensions. 
The optical potential $V_{\rm opt}(\vec{x},t)$ is the sum of a static, periodic part corresponding to the optical lattice $V_{\rm latt}(\vec{x})=V\sum_i\sin^2(kx_i)$ and a fluctuating contribution $V_{\rm noise}(\vec{x},t)$. In the case of global intensity noise we have $V_{\rm noise}(\vec{x},t)=\epsilon(t)V_{\rm latt}(\vec{x})$, so that 
 \begin{align}V_{\rm opt}(\vec{x},t)=(1+\epsilon(t))V_{\rm latt}(\vec{x}).\end{align}
The optical lattice is periodic at each instant in time and thus we can define a set of instantaneous Bloch states $\{\phi_{\vec{q},\vec{n}}(\vec{x},\epsilon(t))\}$ that diagonalize the single particle part of the Hamiltonian at each time $t$. We expand the field operators in this time-dependent basis $\f{\vec{x}}=\sum_{\vec{q},\vec{n}}\phi_{\vec{q},\vec{n}}(\vec{x},\epsilon(t))b_{\vec{q},\vec{n}}(t)$.
By changing from the instantaneous Bloch basis to the instantaneous Wannier basis one obtains the standard (multi-band) Bose-Hubbard Hamiltonian where the parameters are those of a lattice $V_{\rm opt}(\vec{x},t)$ with depth $V(1+\epsilon(t))$. For small fluctuations we can linearize this about the static value. For example the lowest band hopping parameter at time $t$ is given by $J(V(1+\epsilon(t)))\approx J(V)+\frac{dJ}{dV}\delta V(t)$ (with the notation $\delta V(t)=V\epsilon(t)$). For the lowest band this gives the zeroth order terms give the $H_{BH}(J,U)$, the first order term gives $H_{BH}(dJ/dV,dU/dV)$ in eq. \ref{sde}, but with time-dependent mode operators $b_i(t)$.
To remove the time dependence of the creation and annihilation operators $b_{\vec{q},\vec{n}}(t)$, we perform the unitary transformation to time-independent creation and annihilation operators $b_{\vec{q},\vec{n}}$, which introduces the non-adiabatic term $H_{na}$.
\begin{align}
H_{na}=-i\sum_{\vec{q}; \vec{n}\neq \vec{m}}\braket{ \phi_{\vec{q},\vec{n}}(\epsilon(t))}{\dot{\phi}_{\vec{q},\vec{m}}(\epsilon(t))}b_{\vec{q},\vec{n}}^{\dag}b_{\vec{q},\vec{m}}
\end{align}

We note that this term is diagonal in the quasi-momentum since the fluctuations have the same period as the static lattice. Further we note that the terms with $\vec{n}=\vec{m}$ vanish identically. This is readily seen by writing
\begin{align}
\ket{\dot{\phi}_{\vec{q},\vec{n}}(\epsilon(t))}&=\sum_{\vec{m}\neq \vec{n}}\ket{\phi_{q,m}(\epsilon)}\frac{\bra{\phi_{\vec{q},\vec{m}}(\epsilon)}V_{\rm latt}(\vec{x})\ket{{\phi}_{\vec{q},\vec{n}}(\epsilon)}}{\varepsilon_{\vec{q},\vec{n}}-\varepsilon_{\vec{q},\vec{m}}}\dot{\epsilon}(t)
\end{align}

Therefore $H_{na}$ does not couple states within the same band, instead generating only transitions between bands. If the noise has a frequency cutoff well below the separation of the Bloch-bands, this term will be non-resonant and will not drive any transitions on the timescales of interest. In particular we can then restrict the dynamics to the lowest band dropping the non-adiabatic term.

The same argument holds if one starts out with the noise resilient setup. The starting point here is the Hamiltonian
\begin{align}
H&=\sum_{\sigma=g,h}\int d^3x\, \Fd{\vec{x}}{\sigma}\lr{-\frac{\nabla^2}{2m} +V^{(\sigma)}(\vec{x},t)}\F{\vec{x}}{\sigma}\no\\
&-\int d^3x\,\lr{ \Delta_0\Fd{\vec{x}}{h}\F{\vec{x}}{h}+\frac{\Omega_{gh}}{2}\lr{\Fd{\vec{x}}{h}\F{\vec{x}}{g}+\textrm{h.c.}}}\no\\
&+\sum_{\sigma,\sigma'}\frac{g_{\sigma,\sigma'}}{2}\int d^3x\, \Fd{\vec{x}}{\sigma}\Fd{\vec{x}}{\sigma'}\F{\vec{x}}{\sigma'}\F{\vec{x}}{\sigma}.
\end{align}
Again the eigenstates of the single particle part are Bloch states, which now have nonzero amplitudes for both internal states. Expanding the field operator in this set of eigenstates on can repeat all arguments form before to obtain the result, that the non adiabatic term do not cause transitions within an (effective) band.

\section{Details of the noise-resilient setup}

To analyse the noise-resilient setup, and determine parameters to ensure that the required condition is fulfilled, we start from the single particle Hamiltonian $H^{(1)}$ describing a single atom of mass $m$ in the three dimensional isotropic optical lattice $V^{(g)}(\vec{x})=V\sum_{i=1}^3\sin^2(kx_i)$ if it is in the state $\ket{g}$ and in the lattice $V^{(h)}(\vec{x})=-V^{(g)}(\vec{x})$ if it is in the state $\ket{h}$,
\begin{align}
H^{(1)}&=&\frac{\vec{p}^2}{2m}+V^{(g)}(\vec{x})\ket{g}\bra{g}+V^{(h)}(\vec{x})\ket{h}\bra{h}\nonumber\\
& &-\Delta_0\ket{h}\bra{h}-\frac{\Omega_{gh}}{2}\lr{\ket{h}\bra{g}+\ket{g}\bra{h}}.\label{eq:Noise2}
\end{align}
Here, $\Omega_{gh}$ is the coupling strength, and $\Delta_0$ the detuning between the states, where the parameters are taken for the bare atom without the lattice. In the present setup it is convenient also to introduce the detuning $\Delta$ of the two lowest bands in the lattice $\Delta=\Delta_0+3V$. Note that it is not essential for the lattice depths to be identical.

In the uncoupled lattice ($\Omega_{gh}=0$) the eigenstates are Bloch states with quasi-momentum $\vec{q}$ in band $\vec{n}$ and internal state $g$ (denoted by $\ket{\phi_{\vec{q},\vec{n}}^{(g)}}\ket{g}$) or internal state $h$  (denoted by $\ket{\phi_{\vec{q},\vec{n}}^{(h)}}\ket{h}$). The corresponding eigenenergies $\varepsilon_{\vec{q},\vec{n}}^{(g)}$ and $\varepsilon_{\vec{q},\vec{n}}^{(h)}$ differ by $\Delta$. The Bloch states for the $h$ and the $g$ lattice are shifted by half a lattice constant with respect to each other [$\phi_{\vec{q},\vec{n}}^{(h)}(\vec{x})=\phi_{\vec{q},\vec{n}}^{(g)}(\vec{x}-\sum_{i=1}^3\vec{e}_ia/2)$]. From these Bloch states and energies one can construct the Wannier-functions $w_{\vec{n}}^{(g)}(\vec{x})$ and their onsite energies $\varepsilon_{\vec{n}}^{(g)}$ in the standard way \cite{toolbox}. 

For finite $\Omega_{gh}$ the two internal states are coupled. As a consequence of the periodicity only states with the same quasi-momentum couple. In the limit of large detuning $|\Delta|\gg|\Omega_{gh}\braket{\phi_{\vec{q},\vec{n}}^{(h)}}{\phi_{\vec{q},\vec{0}}^{(g)}}|$ we can eliminate the $h$-lattice adiabatically and obtain an effective Hamiltonian for the lowest band in the $g$-lattice $H_{\rm{eff}}^{(1)}=\sum_{\vec{q}}\varepsilon_{\vec{q},\vec{0};\rm{eff}}^{(g)}\ket{\phi_{\vec{q},\vec{0}}^{(g)}}\bra{\phi_{\vec{q},\vec{0}}^{(g)}}$ where the effective  energy band is modified to be
\begin{align}
\varepsilon_{\vec{q},\vec{0};\rm{eff}}^{(g)}=\varepsilon_{\vec{q},\vec{0}}^{(g)}-\sum_{\vec{n}}\frac{|\Omega_{gh}|^2}{4}\frac{|\braket{\phi_{\vec{q},\vec{n}}^{(h)}}{\phi_{\vec{q},\vec{0}}^{(g)}}|^2}{\varepsilon_{\vec{q},\vec{n}}^{(g)}-\Delta-\varepsilon_{\vec{q},\vec{0}}^{(g)}}.\label{eq:Noise9}
\end{align}
This leads to an effective hopping rate
\begin{align}
J_{\rm eff}&=\lr{\frac{a}{2\pi}}^3\iiint\limits_{|q_i|<k} d^3q\,\cos(q_1 a)\varepsilon_{\vec{q},\vec{0};\rm{eff}}^{(g)}.\label{eq:Noise10}
\end{align}
For deep lattices one can show that this is approximately equal to 
\begin{align}
J_{\rm eff}&\approx J+\sum_{\vec{n}} (-1)^{n_1}\frac{|\Omega_{gh}|^2|C_{\vec{n}}|^2}{\varepsilon_{\vec{n}}^{(g)}-\Delta-\varepsilon_{\vec{0}}^{(g)}},\\
C_{\vec{n}}&=\int d^3 x\, w_{\vec{0}}^{(g)}(\vec{x})w_{\vec{n}}^{(g)}\lr{\vec{x}-\sum_{i=1}^3\vec{e}_ia/2}.\label{eq:Noise12}
\end{align}

In addition to $J$ also the interactions are modified due to the admixture of the $h$-lattice. These depend in general on the scattering lengths $a_{\alpha,\beta}$ for collisions of an atom in state $\ket{\alpha}$ with an atom in state $\ket{\beta}$ ($\alpha,\beta\in\{g,h\}$). However to lowest order we can neglect the interactions of the population in state $h$ with the ones in $g$ because these atoms are located in wells separated by half a lattice constant. Interactions between atoms in the state $h$ can be neglected because they are of fourth order in $\Omega_{gh}C_{\vec{n}}/\Delta$. We find in second order perturbation theory, taking into account only the Wannier functions in the lowest band in the two lattices, (which are the closest in energy) expressions for the effective Bose Hubbard parameters
\begin{align}
J_{\rm eff}&\approx J-\frac{|\Omega_{gh}|^2|C_{\vec{0}}|^2}{\Delta}\\
U_{\rm eff}&\approx U\lr{1-8\frac{|\Omega_{gh}|^2|C_{\vec{0}}|^2}{\Delta^2}},\label{eq:Noise11}
\end{align}
Thus the dependence of the effective hopping rate $J_{\rm eff}$ on $V$ depends on the choice of $\Omega_{gh}$ and $\Delta$.
Therefore we can modify the relative change of the effective hopping and interaction rates via the coupling $\Omega_{gh}$ and the detuning $\Delta$, which is needed to minimize heating in the lowest band and eventually reach the sweet spot, where the effects of noise vanish. 
Now, noting that $d\Delta/dV=3$ and neglecting the change of $C_{\vec{0}}$ with $V$, the relation that $\Omega_{gh}$ and $\Delta$ must satisfy in order to produce a sweet spot for the effective Hamiltonian is
\begin{align}
\Omega_{gh}&=-\sqrt{\frac{1}{3}\lr{\frac{1}{J}\frac{dJ}{dV}-\frac{1}{U}\frac{dU}{dV}}\frac{J\Delta}{16J-\Delta}}\frac{\Delta}{C_{\vec{0}}}\nonumber\\
&\approx-\sqrt{\frac{J}{3}\lr{\frac{1}{U}\frac{dU}{dV}-\frac{1}{J}\frac{dJ}{dV}}}\frac{\Delta}{C_{\vec{0}}}
\end{align}
(Here all quantities on the right hand side refer to parameters of the uncoupled lattice).

For all these expressions to hold the condition $|\Omega_{gh}C_n|\ll |\Delta|$ needs to be satisfied. A more careful analysis (not relying on this condition) includes the full band structure which can readily be obtained from the diagonalization of the Hamiltonian \eqref{eq:Noise2}. It is such a calculation that is plotted in Fig.~4c.

\end{document}